\documentclass{article}
\usepackage{iclr2027_conference,times}
\iclrfinalcopy
\IfFileExists{pcrr8t.tfm}{}{}

\usepackage[utf8]{inputenc}
\usepackage[T1]{fontenc}
\usepackage[pagebackref,breaklinks,colorlinks,citecolor=blue,linkcolor=red]{hyperref}

\usepackage{url}
\usepackage{booktabs}
\usepackage{amsfonts}
\usepackage{amssymb}
\usepackage{amsmath}
\usepackage{nicefrac}
\usepackage{microtype}
\usepackage{booktabs}
\usepackage[table]{xcolor}
\usepackage{graphicx}
\usepackage{algorithm}
\usepackage{algorithmic}
\usepackage{multirow}
\usepackage{enumitem}
\usepackage{subcaption}
\usepackage{tikz}
\usetikzlibrary{arrows.meta,calc,positioning}

\newcommand{\circled}[1]{\tikz[baseline=(char.base)]{
  \node[shape=circle,draw,inner sep=1pt] (char) {\small #1};}}

\title{\ours{}: Do Agent Skills Actually Help in Real-World Software Engineering?}
\author{Tingxu Han \footnotemark[1]\hspace{0.5em}\footnotemark[2] \\
  Nanjing University, Mohamed bin Zayed University of Artificial Intelligence \\
  \texttt{txhan@smail.nju.edu.cn} 
  \AND
  Yi Zhang \footnotemark[1]\\
  South China University of Technology \\
  \texttt{202330580551@mail.scut.edu.cn} 
  \And
  Wei Song \\
  Griffith University \\
  \texttt{w.song@griffith.edu.au} 
  \And
  Ye Shang \\
  Nanjing University \\
  \texttt{yeshang@smail.nju.edu.cn} 
  \And
  Chunrong Fang \footnotemark[3]\\
  Nanjing University \\
  \texttt{fangchunrong@nju.edu.cn}
  \And
  Zhenyu Chen \\
  Nanjing University \\
  \texttt{zychen@nju.edu.cn}
  \And
  Youcheng Sun \\
  Mohamed bin Zayed University of Artificial Intelligence \\
  \texttt{youcheng.sun@mbzuai.ac.ae}
  \AND
  Lijie Hu \footnotemark[3]\\
  Mohamed bin Zayed University of Artificial Intelligence \\
  \texttt{lijie.hu@mbzuai.ac.ae} 
}

\newcommand{\ours}{SWE-Skills-Bench}

\begin{document}
\maketitle
\lhead{}
{
\renewcommand{\thefootnote}{\fnsymbol{footnote}}
\footnotetext[1]{Equal Contribution.}
\footnotetext[2]{ Work done during a research visit at MBZUAI.}
\footnotetext[3]{Corresponding Author.}
}
\begin{abstract}
Agent skills, structured procedural knowledge packages injected at inference time, are increasingly used to augment LLM agents on software engineering tasks.
However, their real utility in end-to-end development settings remains unclear, and no standard procedure tests an individual skill before installation.
We present \textbf{\ours{}}, a skill-centric, requirement-driven benchmark that measures how much individual agent skills improve performance on real-world software engineering (SWE) tasks.
We start from 98 public SWE skills, each paired with a real GitHub repository pinned at a fixed commit. 
For each skill, we derive repository-specific requirements, create executable verifiers, and build a reproducible environment. 
Under each of four agent--model configurations, we run every task once without the skill and once with it, yielding 3,920 paired evaluations.
For each requirement, we compute a paired delta, i.e., the change in outcome when the skill is added.
We then aggregate these deltas across a skill's ten requirements into a per-skill utility profile, and use a statistical test to classify each skill as Golden (significantly helpful), Ineffective (no significant effect), or Harmful (significantly detrimental).
Our results show that skills provide limited benefits: 92 of 98 skills have no statistically significant effect, and adding skills raises the average pass ratio only from $76.6\%$ to $76.9\%$.
The few helpful skills supply procedures that the agent would otherwise omit (gains of up to $+17.7$ points). In contrast, harmful skills anchor the agent on their own templates and override the task's explicit instructions (losses of up to $-17.0$ points). 
These findings suggest that skill utility is an empirical, configuration-dependent property rather than an intrinsic one, and \ours{} provides a testbed for testing skills before deployment.
Our code is available at \url{https://github.com/ControllableModel/SWE-Skills-Bench}.

\end{abstract}
\section{Introduction}
\label{sec:intro}
\begin{figure}[t]
    \centering
    \includegraphics[width=0.88\columnwidth]{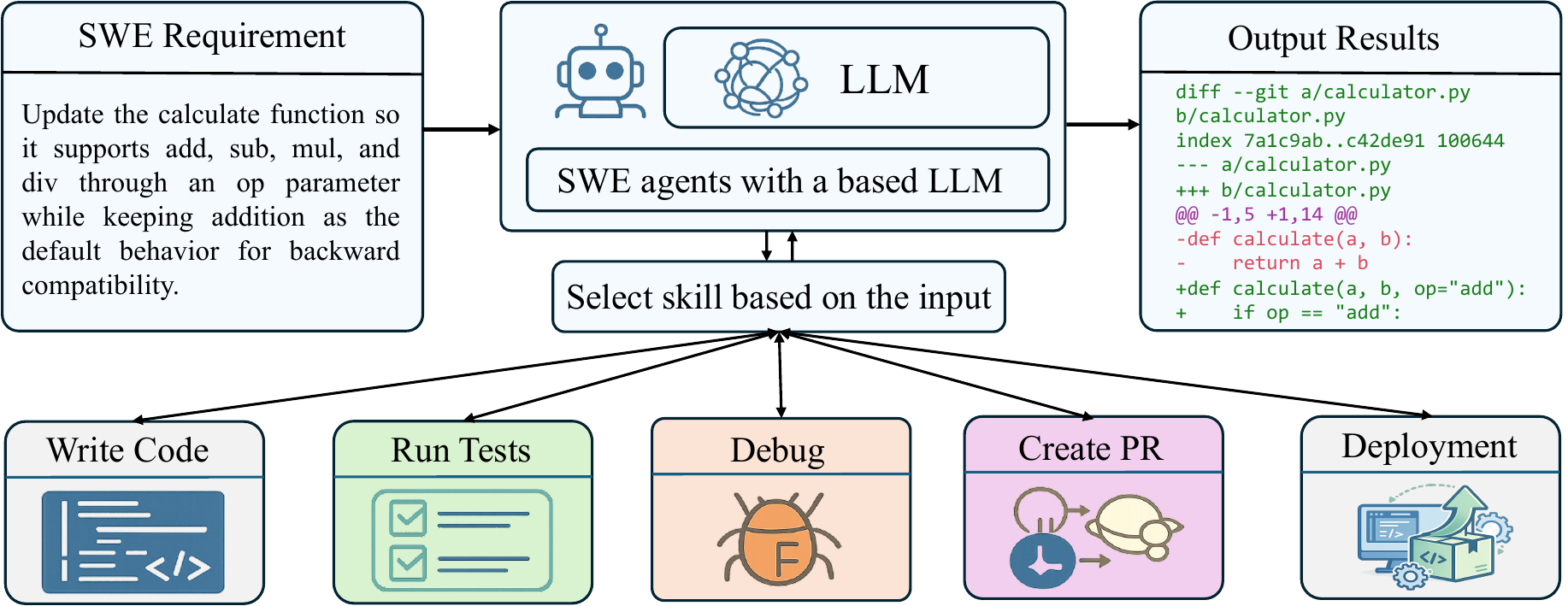}
  \caption{Illustration of how agent skills are used in a software engineering workflow. Given a natural-language requirement, the LLM-based agent selects the most relevant skill from its skill library, including skills such as writing code, running tests, debugging, creating pull requests, and deploying, and injects it into the context window. The agent then executes a series of SWE actions to produce the final software artifacts (such as code) that fulfill the requirement.}
  \vspace{-12pt}
\label{fig:skill-augmentation}
\end{figure}

\vspace{-8pt}
LLM-based agents~\citep{yao2023react,schick2023toolformer} have been increasingly deployed across multiple software engineering (SWE) tasks, from bug fixing~\citep{jimenez2024swebench,zhang2024autocoderover,wang2025openhands,yang2025swesmith} to infrastructure management~\citep{yang2024sweagent,song2025help}.
Agent skills are structured markdown packages that encode procedural knowledge, code templates, and domain conventions for consumption by LLM-based agents~\citep{anthropic2025skills,fang2025memp,wang2023voyager,wang2024survey,xu2026agentskills,wu2026agentskillsmemory}.
At inference time, a skill is injected into the agent's context window as a reference document.
Unlike fine-tuning or retrieval-augmented generation, skills require neither model modification nor an external retrieval pipeline. \autoref{fig:skill-augmentation} illustrates how an agent uses skills to solve a software engineering task.
Public skill ecosystems have grown rapidly. 
Skills.sh, a popular skill website, tracked over 90K installable skills in early 2026~\citep{virtualuncle2026skillssh} and more than 1.55M as of September 2026    ~\citep{skillssh2026}.

This rapid proliferation has outpaced quality assurance.
A public skill can be installed and reused immediately, yet there is no standard way to test whether it actually helps.
Simply observing whether an agent completes a task is not enough, because the outcome mixes together the capability of the foundation model, the behavior of the agent harness, the difficulty of the task, and the contribution of the skill.
Isolating the skill's contribution requires comparing the same task with and without the skill.
Existing benchmarks support such skill-level comparison only partially.
Terminal-Bench~\citep{merrill2026terminalbench} evaluates realistic command-line tasks, HumanEval~\citep{chen2021humaneval}, DS-1000~\citep{lai2022ds1000}, BigCodeBench~\citep{zhuo2025bigcodebench}, and LiveCodeBench~\citep{jain2025livecodebench} evaluate code generation, and agent benchmarks such as AgentBench, $\tau$-Bench, AppWorld, and MLE-Bench~\citep{liu2024agentbench,yao2025taubench,trivedi2024appworld,chan2025mlebench} evaluate tool use and end-to-end task completion.
None of these treats a skill as the intervention under test.
The closest work is SkillsBench~\citep{li2026skillsbench}, which evaluates agents with and without skills on 87 expert-contributed tasks across eight domains, using containerized environments and deterministic verifiers.
However, SkillsBench is \emph{task-centric} but not \emph{skill-centric}: each evaluation starts from a fixed, expert-contributed task and asks whether providing curated skills for that task improves agent performance.
Furthermore, SkillsBench is not specifically designed for software engineering: software engineering is only one of its eight domains, and its primary goal is to measure broad cross-domain skill efficacy rather than requirement satisfaction in real-world development workflows.

A principled benchmark for SWE skill utility must answer a deceptively simple question: \emph{Does the skill help the agent satisfy the requirements it is intended to support?}
Software engineering is inherently requirement-driven~\citep{sommerville2015software,zave1997classification,pohl2010requirements,nuseibeh2000requirements,boehm1981software}: a task succeeds when every acceptance criterion stated in its specification is met, and unit tests serve as executable encodings of those criteria.
We therefore reverse-engineer repository-specific requirements from each skill's claimed procedures, then systematically translate their acceptance criteria into deterministic verifiers.
Building on this principle, we present \textbf{\ours{}}, a skill-centric framework for constructing and conducting automated evaluations of software engineering skills.
We manually curate 98 public SWE skills and pair each with an authentic GitHub repository pinned at a fixed commit.
For every skill--repository pair, our pipeline (i) reverse-engineers ten skill-triggering requirements, (ii) synthesizes requirement-driven executable verifiers, (iii) constructs a reproducible Docker environment for the target repository, and (iv) conducts matched skill-free and skill-assisted executions across agent--model configurations.
Because the paired arms differ only in whether the skill is available, the resulting change in pass ratio can be attributed to the tested skill under that configuration.

Our main contributions are as follows:
\begin{itemize}[leftmargin=*]
    \item \textbf{Automated skill-centric evaluation.} We formulate skill evaluation as an executable, requirement-driven procedure. We then build an end-to-end pipeline that takes a curated skill--repository pair as input, generates requirements, deterministic verifiers, and a containerized environment, and outputs a utility profile for the skill.
    \item \textbf{Empirical findings.}\circled{1}~Skills do not provide a universal performance boost: 92 of 98 skills are Ineffective, and the mean pass ratio changes only from $76.6\%$ without skills to $76.9\%$ with skills ($+0.3$ points).
\circled{2}~Four skills are Golden, and they help by supplying procedures the agent would otherwise omit. For example, \texttt{viper-configuration} ($+17.7$ points) leads the agent to use the \texttt{spf13/viper} library that the requirements expect. Without the skill, half of the runs never add the library and instead write their own configuration loader.
\circled{3}~Skills can also hurt: \texttt{asyncio-patterns} lowers the pass ratio by $17.0$ points on average, and trajectories show a skill's concrete templates overriding the task's explicit instructions, for example by relocating the required files.
\circled{4}~Skill utility is configuration-dependent. 
Skill-level deltas are nearly uncorrelated across configurations. 40 of 98 skills change sign between two harnesses running the same model.
These results establish that SWE skill utility is highly domain-specific, configuration-dependent, and context-dependent, favoring targeted skill design over blanket adoption.
\end{itemize}
\section{Related Benchmarks \& Datasets}
\label{sec:related}
\vspace{-8pt}

We organize related work into two threads: SWE- and Skill-related benchmarks.
Generally, SWE-related benchmarks do not include skills in their evaluations, whereas skill-related benchmarks do not focus specifically on software engineering tasks.
To the best of our knowledge, we are the first skill-centric benchmark for software engineering.
\autoref{tab:comparison} summarizes the key differences.
\begin{table}[!ht]
\centering
\caption{Comparison of \ours{} with existing benchmarks. ``Skill Cond.'' indicates whether the benchmark includes agent skills. ``Det. Verifier'' indicates whether deterministic (non-LLM) verification is included. ``SWE-Focused'' indicates whether the benchmark is specifically designed for software engineering tasks.}
\label{tab:comparison}
\begin{tabular}{lccccc}
\toprule
\textbf{Benchmark} & \textbf{Size} & \textbf{Skill Cond.} & \textbf{Real Projects} & \textbf{Det. Verifier} & \textbf{SWE-Focused} \\
\cmidrule(lr){1-6}
SWE-Bench Verified  & 500 & None & Yes & Yes & Yes \\
TerminalBench  & 89 & None & Yes & Yes & Yes \\
HumanEval  & 164 & None & No & Partial & No \\
SkillsBench  & 87 & Yes & Yes & Yes & Partial \\
\cmidrule(lr){1-6}
\textbf{\ours{}} & \textbf{98} & \textbf{Yes} & \textbf{Yes} & \textbf{Yes} & \textbf{Yes} \\
\bottomrule
\end{tabular}
\end{table}

\textbf{SWE-related Benchmarks.}
This line of work can be further divided into repository-level SWE benchmarks and code-generation benchmarks. The former focus on realistic, project-level software engineering tasks with execution-based verification.
SWE-Bench Verified~\citep{jimenez2024swebench} is a human-validated subset of 500 instances from SWE-Bench, drawn from 12 Python repositories and evaluated via fail-to-pass tests.
TerminalBench~\citep{merrill2026terminalbench} evaluates agents on 89 realistic CLI tasks in containerized environments. However, these benchmarks do not isolate the marginal benefit of injecting procedural skill documents.
Code generation benchmarks, in contrast, mainly evaluate models on self-contained coding problems (often algorithmic or snippet-level) without full project context. HumanEval~\citep{chen2021humaneval} comprises 164 hand-crafted programming challenges at the function level, and therefore does not capture multi-file reasoning, dependency management, or end-to-end SWE workflows.

\textbf{Skills Benchmarks.}
SkillsBench~\citep{li2026skillsbench} takes an important first step toward benchmarking skills as first-class artifacts by comparing agent performance across different skill conditions. Nevertheless, it is not SWE-specific: software engineering forms only a limited subset of its task suite, and the benchmark is not designed around the central success criterion in real-world development—whether explicit requirements are satisfied in repository-grounded workflows. Our work addresses this gap by constructing a requirement-driven benchmark focused exclusively on SWE, pairing each skill with fixed-commit repositories, explicit requirements, and deterministic, execution-based verification.

\section{\ours{}: Requirement-Driven Skill Evaluation}
\label{sec:data}
\vspace{-8pt}
\subsection{Overview and Problem Formulation}
\label{subsec:formulation}
\vspace{-4pt}

\ours{} evaluates a skill by measuring how much it improves the agent's performance on requirements, which are reverse-engineered from the skill content.
The input is a manually curated skill--repository pair $(s,R_s)$, where $R_s$ is an authentic GitHub repository pinned at a fixed commit.
Once the skill and its repository are fixed, we proceed in three steps. First, we reverse-engineer repository-specific requirements from the skill's content. Second, we build executable verifiers for these requirements and a reproducible environment for running them. Third, we run each requirement twice, once with the skill and once without it. \autoref{fig:pipeline} showcases the pipeline.

For each skill $s$, we construct $m=10$ requirements $\mathcal{P}_s=\{P_{s,1},\ldots,P_{s,m}\}$, a containerized environment $E_s$, and a deterministic verifier $V_{s,i}$ for each requirement.
Let $A_c$ denote an agent--model configuration $c$ and let $z\in\{0,1\}$ indicate whether $s$ is unavailable or available, respectively.
The agent produces an artifact
\begin{equation}
    x^{z}_{s,i,c}=A_c(R_s,E_s,P_{s,i};s^{z}),
    \qquad
    s^{z}=\begin{cases}s,&z=1,\\ \varnothing,&z=0,\end{cases}
\end{equation}
which receives the execution-based pass ratio $y^{z}_{s,i,c}=V_{s,i}(x^{z}_{s,i,c})\in[0,1]$, the fraction of executable checks the artifact passes.
Within a matched pair, the repository, requirement, environment, verifier, and agent--model configuration are held fixed; only access to $s$ changes.

\begin{figure}[t]
\centering
\begin{tikzpicture}[
  font=\small,
  io/.style={draw=black!55, line width=0.6pt, rounded corners=4pt, fill=gray!6, inner sep=5pt, align=center, minimum height=0.95cm},
  flow/.style={-{Stealth[length=2.2mm]}, line width=0.9pt, draw=black!60}
]
\def\bw{4.15cm}\def\bh{2.25cm}\def\gap{4.7}
\foreach \i/\x/\col/\ttl/\body in {
  1/0/blue/{1.\ Requirement\\Reverse Engineering}/{skill $s$ + repository $R_s$\\ $\rightarrow$ ten requirements $\{P_{s,i}\}$},
  2/\gap/green/{2.\ Executable Evaluation\\Construction}/{verifier synthesis $P_{s,i}\rightarrow V_{s,i}$\\ AutoDocker $R_s\rightarrow E_s$},
  3/2*\gap/orange/{3.\ Paired Evaluation and\\Statistical Assessment}/{without skill $\leftrightarrow$ with skill\\ $d_{s,i,c}=100\,(y^{1}_{s,i,c}-y^{0}_{s,i,c})$}}{
  \node[minimum width=\bw, minimum height=\bh, inner sep=0pt] (s\i) at (\x,0) {};
  \fill[white, rounded corners=4pt] (s\i.south west) rectangle (s\i.north east);
  \fill[\col!12] ([yshift=-0.9cm]s\i.north west) {[rounded corners=4pt] -- (s\i.north west) -- (s\i.north east)} -- ([yshift=-0.9cm]s\i.north east) -- cycle;
  \draw[\col!60!black, line width=0.6pt] ([yshift=-0.9cm]s\i.north west) -- ([yshift=-0.9cm]s\i.north east);
  \draw[\col!60!black, line width=0.6pt, rounded corners=4pt] (s\i.south west) rectangle (s\i.north east);
  \node[font=\small\bfseries, align=center, text width=\bw-8pt] at ([yshift=-0.45cm]s\i.north) {\ttl};
  \node[align=center, text width=\bw-6pt] at ($([yshift=-0.9cm]s\i.north)!0.5!(s\i.south)$) {\body};
}
\draw[flow] (s1.east) -- (s2.west);
\draw[flow] (s2.east) -- (s3.west);
\node[io, minimum width=\bw, anchor=south] (in) at ([yshift=0.65cm]s1.north)
  {\textbf{Input:} curated skill--repository pair\\ $(s,\;R_s)$ at a fixed commit};
\node[io, minimum width=\bw, anchor=south] (out) at ([yshift=0.65cm]s3.north)
  {\textbf{Output:} per-skill utility profile\\ $\Delta_{s,c}$, $p$-value, class};
\draw[flow] (in.south) -- (s1.north);
\draw[flow] (s3.north) -- (out.south);
\end{tikzpicture}
\caption{Overview of the \ours{} pipeline. Given a curated skill $s$ and a repository $R_s$ at a fixed commit, \ours{} (1)~reverse-engineers ten requirements from the skill, (2)~builds a verifier based on each requirement and a reproducible environment, and (3)~runs each requirement with and without the skill to produce a per-skill utility profile.}
\vspace{-6pt}
\label{fig:pipeline}
\end{figure}

\paragraph{Evaluation inputs.}
We manually curate a skill subset through a three-stage filtering pipeline as evaluation inputs. First, we scan the \texttt{mcpmarket} category leaderboard and select six of the nine core categories that best align with software-engineering workflows and are amenable to unit-test evaluation: Developer Tools, Security \& Testing, API Development, Data Science \& ML, Deployment \& DevOps, and Analytics \& Monitoring. Second, we apply semantic filtering to exclude generative or subjective skills, retaining only those that target concrete SWE actions such as \emph{fix}, \emph{build}, and \emph{develop}. Third, we exclude candidates whose associated repositories are prohibitively large or incur high environment and setup costs. 
This pipeline yields 98 skills distributed across the six categories: API Development (26), Developer Tools (18), Data Science \& ML (17), Deployment \& DevOps (14), Analytics \& Monitoring (12), and Security \& Testing (11).

\subsection{Requirement Reverse Engineering}
\label{subsec:requirements}
\vspace{-4pt}

We give the skill document, the repository metadata, and the relevant project structure to a fixed requirement-engineer prompt.
The prompt produces ten distinct requirements, each designed to exercise a different aspect of the skill.
In contrast to task-centric construction, the skill determines the target capability first, and each requirement is then written to test it. 

Each $i$-th requirement of skill $s$, $P_{s, i}$, follows a standardized template with four fields: (i) \emph{Background}, which supplies repository context; (ii) \emph{Requirement}, which states the intended change; (iii) \emph{File Operations}, which identifies artifacts that may need to be created or modified; and (iv) \emph{Acceptance Criteria}, which specify observable success conditions.
\autoref{fig:requirement_gen_prompt} shows the prompt used to generate the requirements, while \autoref{fig:example_requirement} provides an example of a generated requirement.

\subsection{Executable Evaluation Construction}
\label{subsec:construction}
\vspace{-4pt}
Natural-language requirements become executable evaluation instances only after we construct both a scoring procedure and a reproducible environment.
We implement these two components as requirement-driven verification and AutoDocker environment construction.

\paragraph{Requirement-driven Verification.}
Rather than relying on subjective judgments, we convert every acceptance criterion in the requirement document $P_{s,i}$ into objective, deterministic tests, ensuring that each test outcome is directly traceable to a specific requirement.
We provide $P_{s,i}$ (together with repository metadata such as repo path, language, and available test commands) to a fixed “professional test engineer” prompt template, which instructs the model to (i) enumerate testable behaviors from each acceptance criterion, (ii) instantiate representative and edge-case scenarios, and (iii) encode them into a deterministic \texttt{pytest} test file with strong discriminative power (i.e., tests must run the produced code and verify concrete outputs/structures rather than keyword-level heuristics). The prompt also enforces structural constraints such as a minimum number of test cases and per-test docstrings. 
The prompt template is shown in \autoref{fig:test_gen_prompt}.

\paragraph{AutoDocker environment construction.}
Our AutoDocker module provisions $E_s$ from the pinned repository and its environment specification.
It selects a language-appropriate base image, restores a clean checkout at the designated commit, and either executes explicit setup commands or detects dependency installation from common project manifests and lock files.
The resulting container fixes the workspace, dependencies, CPU and memory limits, and command timeouts used by both conditions.
Before evaluation, pre-existing agent-skill directories are removed so that the assisted condition exposes only the skill under test.
This construction ensures that environmental variation cannot be mistaken for skill utility.

\subsection{Paired Evaluation and Statistical Assessment}
\label{subsec:paired_evaluation}
\vspace{-4pt}
For each pair of skill--requirement--configuration $(s,P_{s,i},c)$, we initialize two copies of the same container state and provide the same requirement to the agent.
In the assisted condition, we invoke skill $s$ explicitly at the start of the prompt through the harness's native command (\texttt{/skill} in Claude Code, \texttt{\$skill} in Codex), which loads the full skill document into the agent's context, whereas in the unassisted condition, the skill is absent.
The requirement-level delta, expressed in points, and the skill-level delta under configuration $c$ are
\begin{equation}
    d_{s,i,c}=100\left(y^{1}_{s,i,c}-y^{0}_{s,i,c}\right),
    \qquad
    \Delta_{s,c}=\frac{1}{m}\sum_{i=1}^{m}d_{s,i,c}.
\end{equation}
Positive $\Delta_{s,c}$ indicates improvement under configuration $c$, whereas negative values indicate interference; the overall skill-level delta $\Delta_s$ averages each requirement's delta across configurations before averaging over requirements.

To decide whether a skill has a reliable effect, we test its ten requirement-level deltas with an exact two-sided paired sign-flip test.
Intuitively, the test asks whether the ten deltas favor one direction more consistently than chance alone would produce.
The $p$-value is the probability of seeing a skill-level delta this large if the skill had no real effect.
We call the resulting summary of a skill its \emph{utility profile}. For each configuration, it contains the requirement-level deltas, the skill-level delta, the $p$-value, and the resulting class.
A skill is \emph{Golden} if its skill-level delta is positive with $p<0.05$, \emph{Harmful} if it is negative with $p<0.05$, and \emph{Ineffective} otherwise. Ineffective means that no directional effect was detected with ten requirements, not that the effect is zero.

\section{Results of \ours{}}
\label{sec:experiments}
\vspace{-8pt}

\subsection{Experimental Setup}
\vspace{-4pt}
We evaluate 98 skills, each with ten requirements, under four agent--model configurations: Codex with GPT-5.4-mini or Kimi-K2.6, and Claude Code~\citep{anthropic2025claudecode} with Kimi-K2.6 or Haiku 4.5.
Each requirement is evaluated with and without the target skill in the containerized environment described in \autoref{subsec:paired_evaluation}.
The evaluation spans 980 distinct requirements, 392 skill configuration profiles (agent--model), and all 3,920 expected skill-free/skill-assisted pairs.
Runs affected by infrastructure failures, such as unreachable model endpoints, rate limiting, or lost containers, are re-executed until a clean run is obtained.
Each pass ratio is bound to the trajectory of the run that produced it.

\begin{figure}[!htbp]
\centering
\includegraphics[width=\linewidth]{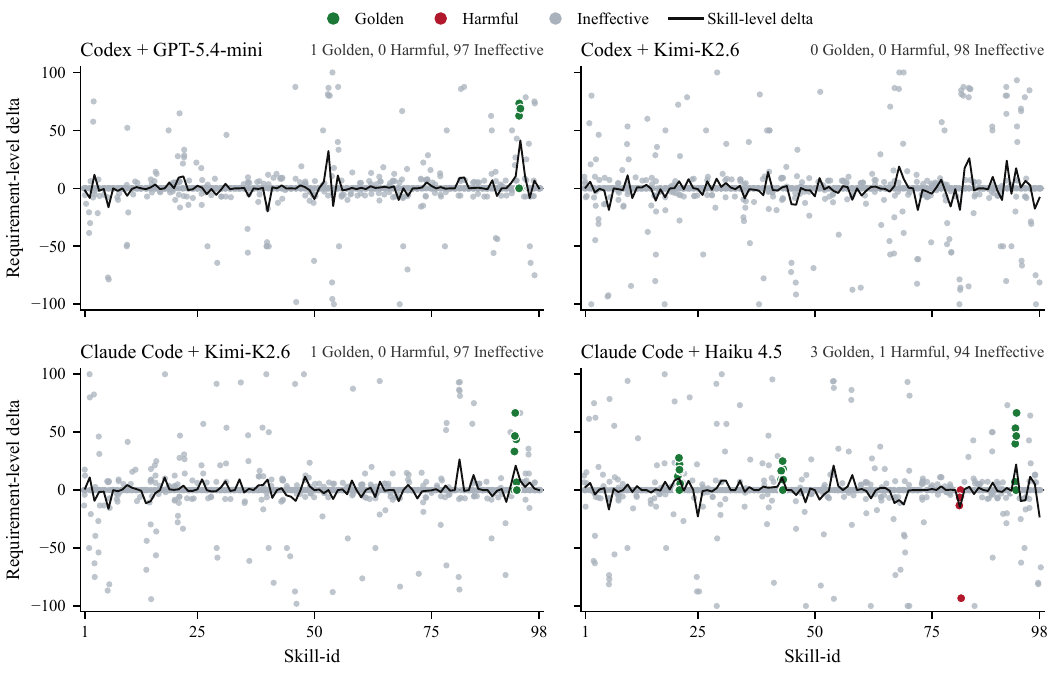}
\caption{Requirement-level deltas for every skill under each configuration: one point per skill-free/skill-assisted pair (980 per panel), skills in alphabetical order with the same index in all panels. The line connects skill-level deltas; colors give the configuration-specific classification (green Golden, red Harmful, gray Ineffective).}
\vspace{-6pt}
\label{fig:s4_requirement_scatter}
\end{figure}

\subsection{Evaluation Results}
\vspace{-4pt}
For each agent--model configuration, the evaluation assigns every skill a utility profile.
\autoref{fig:s4_requirement_scatter} shows the requirement-level deltas for all skills, and \autoref{tab:results} compares average performance with and without skills.
Skill-level deltas are close to zero.
In every configuration, the mean pass ratio changes by less than one point.
Requirement-level deltas are occasionally large ($\pm 50$ points or more), but gains and losses occur at similar rates and cancel out within most skills.

\textbf{Finding 1: Most skills show no detectable directional effect.}
Under our classification protocol, 92 of the 98 skills (93.9\%) are classified as Ineffective, compared with four Golden and two Harmful skills. Skill access increases the aggregate pass ratio only from 76.59\% to 76.93\%, and the difference remains below 1\% in every agent--model configuration. As shown in \autoref{fig:s4_requirement_scatter}, skill-level deltas are therefore small for most skills and directionally consistent for only a few.

At the requirement level, 2,711 of the 3,920 pairs (69.2\%) have identical verifier outcomes with and without the skill. In 1,313 pairs, both runs pass every check; in the remaining 1,398, both pass the same subset of checks. These unchanged outcomes are not explained solely by ceiling effects. Among the skills classified as Ineffective, 35 have unassisted pass ratios below 80\%, yet for 20 of them, the absolute difference between the assisted and unassisted pass ratios is below 0.02.

The small aggregate effect nevertheless masks occasional large changes. In 240 pairs (6.1\%), the pass ratios under the two conditions differ by at least 0.5. In approximately half of these pairs, one run passes fewer than 20\% of the checks while the other passes more than 80\%. Thus, the near-zero average reflects a combination of many unchanged outcomes and a small number of large gains and losses, rather than uniformly small effects across all requirements.

\begin{table}[!tbp]
\centering
\caption{Mean pass ratio over paired with/without runs. Each configuration contributes 980 pairs, and the last row pools all 3,920 pairs. 
Pass ratios are percentages, and $\Delta$ is the mean requirement-level delta over the row's pairs.
Green and red mark observed direction only.}
\label{tab:results}
\begin{tabular}{@{}llrrr@{}}
\toprule
\textbf{Agent} & \textbf{Model} & \textbf{Without Skill} & \textbf{With Skill} & $\boldsymbol{\Delta}$ \textbf{(points)} \\
\midrule
Codex & GPT-5.4-mini & 75.98 & 76.74 & \cellcolor{green!15}$+0.76$ \\
Codex & Kimi-K2.6 & 77.60 & 77.32 & \cellcolor{red!15}$-0.29$ \\
Claude Code & Kimi-K2.6 & 78.53 & 79.18 & \cellcolor{green!15}$+0.64$ \\
Claude Code & Haiku 4.5 & 74.24 & 74.50 & \cellcolor{green!15}$+0.26$ \\
\midrule
\multicolumn{2}{@{}l}{\textbf{All pairs}} & \textbf{76.59} & \textbf{76.93} & \cellcolor{green!15}$\mathbf{+0.34}$ \\
\bottomrule
\end{tabular}
\end{table}

\textbf{Finding 2: Only a few skills improve performance overall.}
Four skills receive an overall Golden classification under our protocol: \texttt{viper-configuration}, \texttt{go-testify-patterns}, \texttt{junit5-testing}, and \texttt{fastapi-patterns} (\autoref{tab:skill_profiles}). Among them, \texttt{viper-configuration} produces the largest improvement, increasing its aggregate pass ratio from 59.61\% to 77.30\%. Its effect is positive in all four agent--model configurations and in nine of the ten requirements. 

\begin{figure}[tbp]
\centering
\includegraphics[width=\linewidth]{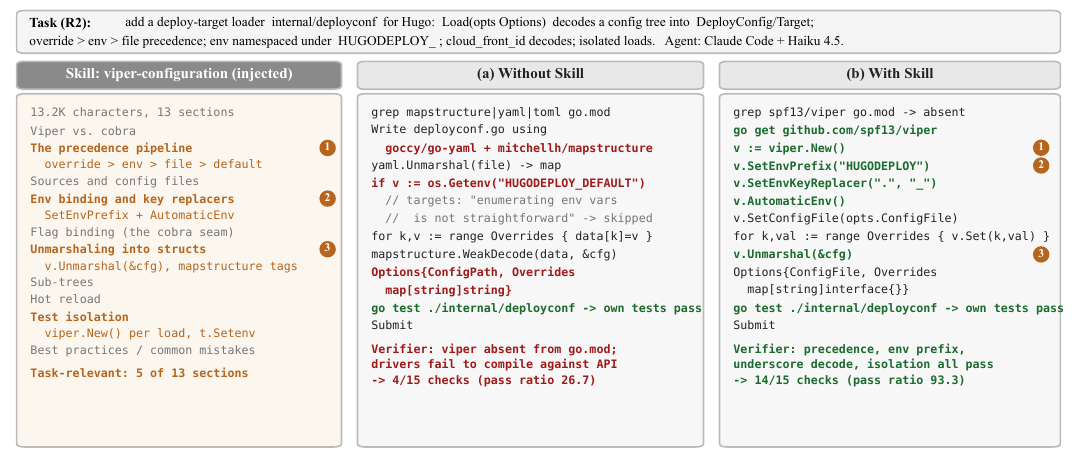}
\caption{
Case study of the \texttt{viper-configuration} skill on a Hugo deployment task (R2).
The agent must implement a configuration loader that respects a specific precedence order (override > env > file) and handles environment variable namespacing.
Without Skill (a), the agent fails to identify the standard library (viper) and instead writes a fragile, hand-rolled YAML parser. This results in a low pass ratio (26.7\%) as it misses critical features like automatic env binding.
With Skill (b), guided by the skill's procedural knowledge, the agent correctly imports viper and implements the required logic (isolated instance, env prefix, unmarshalling), achieving a high pass ratio (93.3\%).}
\vspace{-6pt}
\label{fig:finding2_case}
\end{figure}

The \texttt{viper-configuration} case in \autoref{fig:finding2_case} illustrates one way in which a skill can help. Its requirements exercise configuration workflows involving namespaced environment variables, override precedence, typed decoding, and isolation between configuration loads. Across the ten requirements and four configurations, all 40 assisted runs integrate \texttt{spf13/viper}, whereas only 20 of the 40 unassisted runs do so. Assisted agents also more consistently follow the corresponding Viper workflow. In contrast, unassisted agents often construct ad hoc configuration loaders that pass their self-authored tests but fail verifier checks involving compilation, precedence, decoding, or isolation. The skill therefore provides a procedural mapping from the required configuration behavior to concrete library operations that the unassisted agent frequently omits.

The remaining Golden skills show smaller gains, and their effects vary across requirements and configurations. The Viper case should therefore be viewed as an illustrative example rather than a general explanation for all positive skill-level deltas.

\begin{figure}[!thbp]
\centering
\includegraphics[width=\linewidth]{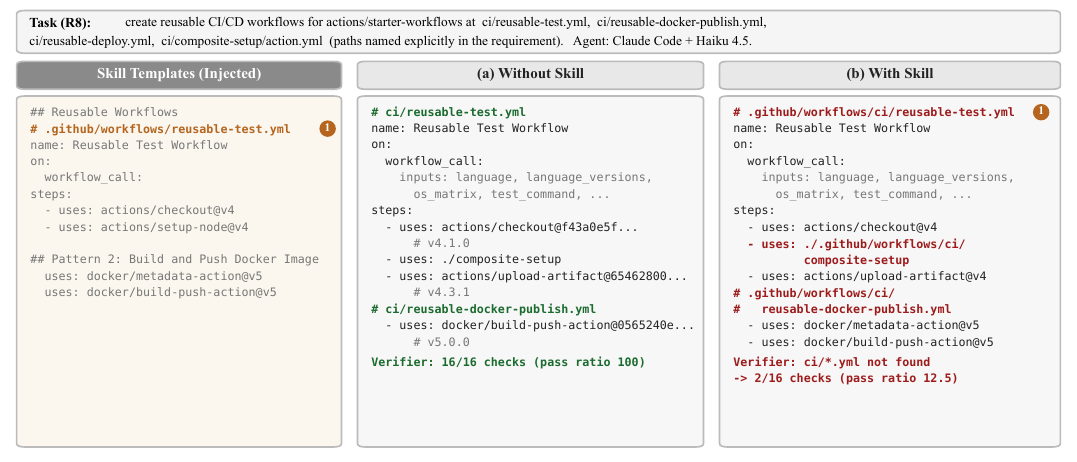}
\caption{
Case study of context interference from the \texttt{github-actions-templates} skill (Claude Code with Haiku 4.5, R8, $100\rightarrow12.5$).
The task requires four reusable workflows at explicitly named paths under \texttt{ci/}.
\textbf{(a)}~Without the skill, the agent writes the files where the requirement specifies and passes all 16 checks.
\textbf{(b)}~With the skill, the agent adopts the location named in the skill's template and writes the files under \texttt{.github/workflows/ci/}.
}
\vspace{-7pt}
\label{fig:finding3_case}
\end{figure}

\textbf{Finding 3: Skills can actively degrade performance through context interference.}
Two skills receive a Harmful label overall: \texttt{grafana-dashboards} ($-2.96$ points) loses a few points on seven of its ten requirements, and \texttt{asyncio-patterns} ($-16.96$ points) falls from a pass ratio of 83.20 to 66.24, with negative deltas on eight of ten requirements and in every configuration ($-16.12$ to $-18.61$ points).
Regressions of this kind also appear within single configurations.
\texttt{github-actions-templates} is Ineffective overall ($-5.14$ points) but loses $22.78$ points under Claude Code with Haiku 4.5, where three of its ten requirements collapse from $75-100$ to $0-12.5$.
These regressions stem from the mismatch between the \emph{holistic} scope of a skill and the \emph{focused} requirements of individual tasks.
Each skill is authored as a comprehensive reference for its technical domain, encoding best practices that span project layout and configuration templates.
When a task exercises only a narrow slice of this knowledge, the surplus context can interfere with the agent's work.
\autoref{fig:finding3_case} illustrates an example of skill \texttt{github-actions-templates}.
The task asks the agent to add four reusable CI/CD workflows to \texttt{actions/starter-workflows} at explicitly named paths under \texttt{ci/}.
The skill packages workflow templates covering testing, Docker build-and-push, Kubernetes deployment, matrix builds, reusable workflows, security scanning, and deployment approvals.
Among them, the \emph{Reusable Workflows} section contains a template of exactly the kind of workflow the task requires.
But the file location is different.
The template's first line places it at \texttt{.github/workflows/reusable-test.yml}, the directory in which GitHub executes a repository's own workflows, whereas the task names \texttt{ci/}.
Because the template matches the task so closely, the agent copies its details, including this location, instead of following the task's instructions.

Without the skill, the agent follows the requirement.
It writes the four files under \texttt{ci/}, references the composite action as \texttt{./composite-setup}, and passes all 16 verifier checks.
With the skill, the agent adopts the template's location instead and writes the files under \texttt{.github/workflows/ci/}, adjusting even the composite-action reference to the new path.
The verifier reads the files from the paths the requirement specifies, finds none of them, and 14 of its 16 checks fail (pass ratio $100\rightarrow12.5$).
The same displacement recurs on the two other collapsed requirements (R4 and R5), whose files also migrate into \texttt{.github/workflows/}, which accounts for the entire loss of this skill under Haiku 4.5.
Misplaced deliverables are a recurring signature of harmful injection.
Among the 118 requirement pairs whose delta is $-50$ points or worse, the verifier reports the assisted run's deliverable as missing or misplaced in 39 (33\%), never the unassisted run's.
This explains the seemingly paradoxical outcome.
A skill containing \emph{objectively relevant} content nonetheless \emph{degrades} performance.
The practical implication is that skill design should favor abstract guidance patterns over concrete, opinionated templates with hard-coded paths and values, as the latter risk anchoring the agent on specifics that the task overrides.

\begin{figure}[!tbp]
\centering
\includegraphics[width=\linewidth]{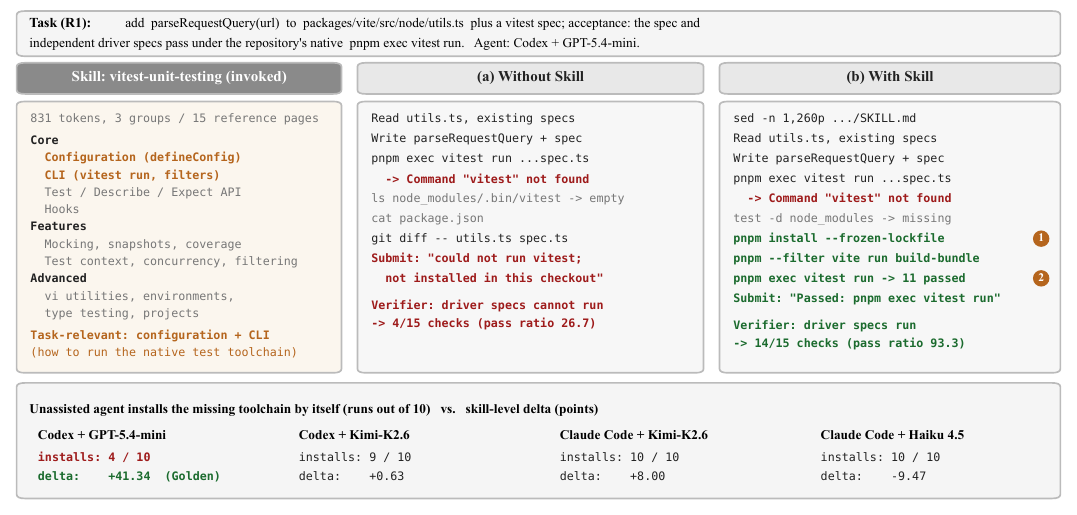}
\caption{
Case study of \texttt{vitest-unit-testing} (Codex with GPT-5.4-mini, R1, $26.7\rightarrow93.3$).
\textbf{(a)}~Without the skill, the agent hits \texttt{Command "vitest" not found} and submits untested code.
\textbf{(b)}~With the skill, it \protect\circled{1}~installs the dependencies and \protect\circled{2}~runs the tests to green.
However, other configurations already install the toolchain on their own, so the skill adds little.
}
\vspace{-6pt}
\label{fig:finding4_case}
\end{figure}

\textbf{Finding 4: Skill utility depends on the agent--model configuration.}
Skill-level deltas transfer poorly across configurations. Only six skills are positive, and five have negative deltas in all four configurations, while the correlation of skill-level deltas between any two configurations ranges from $-0.11$ to $0.21$. Even when the model is held fixed, the harness matters: with Kimi-K2.6, the observed deltas of 40 of the 98 skills change sign between Codex and Claude Code, as shown in \autoref{fig:s4_kimi_harness}.

The \texttt{vitest-unit-testing} case in \autoref{fig:finding4_case} illustrates why the same skill can have different effects across configurations. The requirements are verified using the repository's native \texttt{vitest} framework, but the project dependencies are not installed in the initial checkout. The agent must therefore install the workspace before it can run the tests. Without the skill, Codex with GPT-5.4-mini installs the required toolchain in only four of the ten requirements and otherwise submits code without running the native test suite. With the skill, it installs the workspace and runs the relevant tests successfully in all ten requirements. The skill consequently receives a Golden classification under this configuration and produces the largest configuration-specific improvement reported in \autoref{tab:skill_profiles}.

The other three configurations already install the missing toolchain in nine or ten of the unassisted runs. The skill therefore provides no consistent additional benefit and remains Ineffective under these configurations. In this case, the benefit comes primarily from changing the agent's setup and verification workflow rather than supplying task-specific testing knowledge. A skill's measured utility is thus the marginal difference between the behavior it induces and the behavior the configuration already exhibits, rather than a fixed property of the skill itself.

\section{Discussion \& Future Directions}
\label{sec:discussion}
\vspace{-8pt}

Our results imply that skill utility is not an intrinsic property of a skill but a marginal effect relative to the default behaviour of an agent--model configuration. A skill helps when it supplies a concrete procedure the unassisted agent would otherwise omit, as with \texttt{viper-configuration} and \texttt{vitest-unit-testing}.
It hurts when its concrete templates override the task's own instructions, as with \texttt{github-actions-templates}, or when broad, poorly matched guidance degrades performance across configurations, as with \texttt{asyncio-patterns}. 
Practically, skill authors should favor compact, operational guidance over hard-coded templates, and users should evaluate a skill with the model, agent, and repositories in which it will be deployed, since performance under one configuration does not prove universal utility.

Some limitations remain. The requirements are synthesized from each skill, so they test its claimed capabilities rather than the natural distribution of engineering requests.
Our requirement-driven verification avoids LLM judging but may miss intended behaviors or favor particular implementations.
Future work will audit verifiers through human review, repeat each evaluation, increase the number of requirements per skill to raise statistical power, and extend the benchmark to additional models, harnesses, and multi-skill settings.

\section{Conclusion}
\label{sec:conclusion}
\vspace{-8pt}

We introduced \ours{}, a skill-centric framework for evaluating the marginal utility of agent skills through repository-grounded requirements, deterministic verifiers, and paired executions. Our findings show that providing a skill does not necessarily improve an agent's performance. Its value depends on whether it supplies procedures missing from the agent's default behaviour, while templates that override the task's own instructions may instead interfere with task execution. Skill utility should therefore be treated as configuration-specific rather than as an intrinsic property of the skill. \ours{} provides a practical foundation for evaluating skills with the models, agent harnesses, and task settings in which they will be deployed.

\vspace{-8pt}

\bibliographystyle{iclr2027_conference}
\bibliography{references}

@inproceedings{jimenez2024swebench,
  title={{SWE}-Bench: Can Language Models Resolve Real-World {GitHub} Issues?},
  author={Jimenez, Carlos E. and Yang, John and Wettig, Alexander and Yao, Shunyu and Pei, Kexin and Press, Ofir and Narasimhan, Karthik},
  booktitle={International Conference on Learning Representations (ICLR)},
  year={2024}
}

@misc{skillssh2026,
  title        = {skills.sh: The Agent Skills Directory},
  author       = {{Vercel}},
  year         = {2026},
  howpublished = {\url{https://skills.sh/}},
  note         = {Accessed: 2026-09-25}
}

@misc{virtualuncle2026skillssh,
  title        = {Agent Skills Marketplace: Skills.sh Guide (2026)},
  author       = {{Virtual Uncle}},
  year         = {2026},
  howpublished = {\url{https://virtualuncle.com/agent-skills-marketplace-skills-sh-2026/}},
  note         = {Accessed: 2026-09-25}
}

@article{li2026skillsbench,
  title={{SkillsBench}: Benchmarking How Well Agent Skills Work Across Diverse Tasks},
  author={Li, Xiangyi and Liu, Yimin and Chen, Wenbo and others},
  journal={arXiv preprint arXiv:2602.12670},
  year={2026}
}

@article{merrill2026terminalbench,
  title={Terminal-Bench: Benchmarking Agents on Hard, Realistic Tasks in Command Line Interfaces},
  author={Merrill, Mike A. and others},
  journal={arXiv preprint arXiv:2601.11868},
  year={2026}
}

@misc{anthropic2025skills,
  title={Equipping Agents for the Real World with Agent Skills},
  author={{Anthropic}},
  year={2025},
  howpublished={Anthropic Engineering Blog},
  note={2025a}
}

@misc{anthropic2025claudecode,
  title={Claude Code: An Agentic Coding Tool},
  author={{Anthropic}},
  year={2025},
  howpublished={GitHub},
  note={2025b}
}

@inproceedings{yang2025swesmith,
  title={{SWE}-Smith: Scaling Data for Software Engineering Agents},
  author={Yang, John and Lieret, Kilian and others},
  booktitle={Advances in Neural Information Processing Systems (NeurIPS)},
  year={2025}
}

@inproceedings{yang2024sweagent,
  title={{SWE}-Agent: Agent-Computer Interfaces Enable Automated Software Engineering},
  author={Yang, John and Prabhakar, Akshara and others},
  booktitle={Advances in Neural Information Processing Systems (NeurIPS)},
  year={2024}
}

@inproceedings{zhuo2025bigcodebench,
  title={{BigCodeBench}: Benchmarking Code Generation with Diverse Function Calls and Complex Instructions},
  author={Zhuo, Terry Yue and others},
  booktitle={International Conference on Learning Representations (ICLR)},
  year={2025}
}

@inproceedings{chan2025mlebench,
  title={{MLE}-Bench: Evaluating Machine Learning Agents on Machine Learning Engineering},
  author={Chan, Jun Shern and others},
  booktitle={International Conference on Learning Representations (ICLR)},
  year={2025}
}

@inproceedings{yao2025taubench,
  title={$\tau$-Bench: A Benchmark for Tool-Agent-User Interaction in Real-World Domains},
  author={Yao, Shunyu and others},
  booktitle={International Conference on Learning Representations (ICLR)},
  year={2025}
}

@inproceedings{trivedi2024appworld,
  title={{AppWorld}: A Controllable World of Apps and People for Benchmarking Interactive Coding Agents},
  author={Trivedi, Harsh and others},
  booktitle={Annual Meeting of the Association for Computational Linguistics (ACL)},
  year={2024}
}

@inproceedings{chen2021humaneval,
  title={Evaluating Large Language Models Trained on Code},
  author={Chen, Mark and others},
  booktitle={arXiv preprint arXiv:2107.03374},
  year={2021}
}

@inproceedings{lai2022ds1000,
  title={{DS}-1000: A Natural and Reliable Benchmark for Data Science Code Generation},
  author={Lai, Yuhang and others},
  booktitle={International Conference on Machine Learning (ICML)},
  year={2022}
}

@article{xu2026agentskills,
  title   = {Agent Skills for Large Language Models: Architecture, Acquisition, Security, and the Path Forward},
  author  = {Xu, Renjun and Yan, Yang},
  journal = {arXiv preprint arXiv:2602.12430},
  year    = {2026}
}

@article{wu2026agentskillsmemory,
  title   = {Agent Skills from the Perspective of Procedural Memory: A Survey},
  author  = {Wu, Yaxiong and Zhang, Yongyue},
  journal = {TechRxiv preprint},
  year    = {2026}
}

@article{wang2024survey,
  title   = {A Survey on Large Language Model based Autonomous Agents},
  author  = {Wang, Lei and Ma, Chen and Feng, Xueyang and Zhang, Zeyu and Yang, Hao and Zhang, Jingsen and Chen, Zhiyuan and Tang, Jiakai and Chen, Xu and Lin, Yankai and Zhao, Wayne Xin and Wei, Zhewei and Wen, Ji-Rong},
  journal = {Frontiers of Computer Science},
  volume  = {18},
  number  = {6},
  pages   = {186345},
  year    = {2024}
}

@article{wang2023voyager,
  title   = {Voyager: An Open-Ended Embodied Agent with Large Language Models},
  author  = {Wang, Guanzhi and Xie, Yuqi and Jiang, Yunfan and Mandlekar, Ajay and Xiao, Chaowei and Zhu, Yuke and Fan, Linxi and Anandkumar, Anima},
  journal = {Transactions on Machine Learning Research},
  year    = {2024}
}

@article{fang2025memp,
  title   = {Mem\textsuperscript{p}: Exploring Agent Procedural Memory},
  author  = {Fang, Runnan and Liang, Yuan and Wang, Xiaobin and Wu, Jialong and Qiao, Shuofei and Xie, Pengjun and Huang, Fei and Chen, Huajun and Zhang, Ningyu},
  journal = {arXiv preprint arXiv:2508.06433},
  year    = {2025}
}

@article{zave1997classification,
  title     = {Classification of Research Efforts in Requirements Engineering},
  author    = {Zave, Pamela},
  journal   = {ACM Computing Surveys},
  volume    = {29},
  number    = {4},
  pages     = {315--321},
  year      = {1997},
  publisher = {ACM}
}

@book{pohl2010requirements,
  title     = {Requirements Engineering: Fundamentals, Principles, and Techniques},
  author    = {Pohl, Klaus},
  year      = {2010},
  publisher = {Springer},
  address   = {Heidelberg}
}

@book{sommerville2015software,
  title     = {Software Engineering},
  author    = {Sommerville, Ian},
  edition   = {10th},
  year      = {2015},
  publisher = {Pearson Education}
}

@book{boehm1981software,
  title     = {Software Engineering Economics},
  author    = {Boehm, Barry W.},
  year      = {1981},
  publisher = {Prentice-Hall},
  address   = {Englewood Cliffs, NJ}
}

@inproceedings{nuseibeh2000requirements,
  title     = {Requirements Engineering: A Roadmap},
  author    = {Nuseibeh, Bashar and Easterbrook, Steve},
  booktitle = {Proceedings of the Conference on the Future of Software Engineering},
  pages     = {35--46},
  year      = {2000},
  publisher = {ACM}
}

@article{song2025help,
  title={Help or Hurdle? Rethinking Model Context Protocol-Augmented Large Language Models},
  author={Song, Wei and Zhong, Haonan and Ding, Ziqi and Xue, Jingling and Li, Yuekang},
  journal={arXiv preprint arXiv:2508.12566},
  year={2025}
}

@inproceedings{yao2023react,
  title={{ReAct}: Synergizing Reasoning and Acting in Language Models},
  author={Yao, Shunyu and Zhao, Jeffrey and Yu, Dian and Du, Nan and Shafran, Izhak and Narasimhan, Karthik R. and Cao, Yuan},
  booktitle={International Conference on Learning Representations (ICLR)},
  year={2023}
}

@inproceedings{schick2023toolformer,
  title={Toolformer: Language Models Can Teach Themselves to Use Tools},
  author={Schick, Timo and Dwivedi-Yu, Jane and Dess{\`\i}, Roberto and Raileanu, Roberta and Lomeli, Maria and Hambro, Eric and Zettlemoyer, Luke and Cancedda, Nicola and Scialom, Thomas},
  booktitle={Advances in Neural Information Processing Systems (NeurIPS)},
  year={2023}
}

@inproceedings{wang2025openhands,
  title={{OpenHands}: An Open Platform for {AI} Software Developers as Generalist Agents},
  author={Wang, Xingyao and Li, Boxuan and Song, Yufan and Xu, Frank F. and Tang, Xiangru and Zhuge, Mingchen and Pan, Jiayi and Song, Yueqi and Li, Bowen and Singh, Jaskirat and Tran, Hoang H. and Li, Fuqiang and Ma, Ren and Zheng, Mingzhang and Qian, Bill and Shao, Yanjun and Muennighoff, Niklas and Zhang, Yizhe and Hou, Binyuan and Sun, Yu and Yin, Zhichao and Li, Ruilin and Ji, Heng and Neubig, Graham},
  booktitle={International Conference on Learning Representations (ICLR)},
  year={2025}
}

@inproceedings{zhang2024autocoderover,
  title={{AutoCodeRover}: Autonomous Program Improvement},
  author={Zhang, Yuntong and Ruan, Haifeng and Fan, Zhiyu and Roychoudhury, Abhik},
  booktitle={Proceedings of the 33rd ACM SIGSOFT International Symposium on Software Testing and Analysis (ISSTA)},
  year={2024}
}

@inproceedings{jain2025livecodebench,
  title={{LiveCodeBench}: Holistic and Contamination Free Evaluation of Large Language Models for Code},
  author={Jain, Naman and Han, King and Gu, Alex and Li, Wen-Ding and Yan, Fanjia and Zhang, Tianjun and Wang, Sida and Solar-Lezama, Armando and Sen, Koushik and Stoica, Ion},
  booktitle={International Conference on Learning Representations (ICLR)},
  year={2025}
}

@inproceedings{liu2024agentbench,
  title={{AgentBench}: Evaluating {LLMs} as Agents},
  author={Liu, Xiao and Yu, Hao and Zhang, Hanchen and Xu, Yifan and Lei, Xuanyu and Lai, Hanyu and Gu, Yu and Ding, Hangliang and Men, Kaiwen and Yang, Kejuan and Zhang, Shudan and Deng, Xiang and Zeng, Aohan and Du, Zhengxiao and Zhang, Chenhui and Shen, Sheng and Zhang, Tianjun and Su, Yu and Sun, Huan and Huang, Minlie and Dong, Yuxiao and Tang, Jie},
  booktitle={International Conference on Learning Representations (ICLR)},
  year={2024}
}
\newpage
\newpage
\appendix
\section{Complete Results and Sensitivity Analyses}
\label{sec:s4_supplement}

\begin{figure}[t]
\centering
\includegraphics[width=0.5\linewidth]{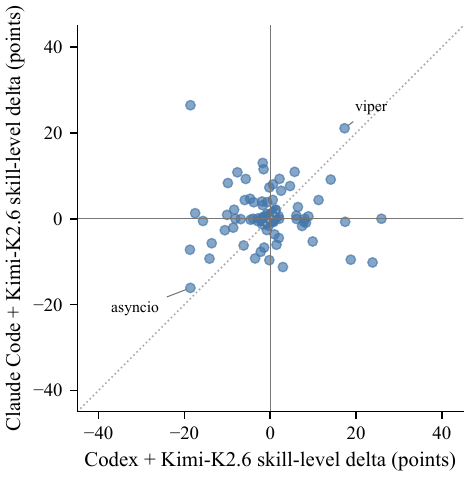}
\caption{Skill-level deltas under the shared Kimi-K2.6 model, with each skill averaged over the same ten requirements in both harnesses. Axes share a scale; the diagonal marks equal effects. The plot describes observed effects, not significant interactions.}
\label{fig:s4_kimi_harness}
\end{figure}

\begin{figure}[t]
\centering
\includegraphics[width=\linewidth]{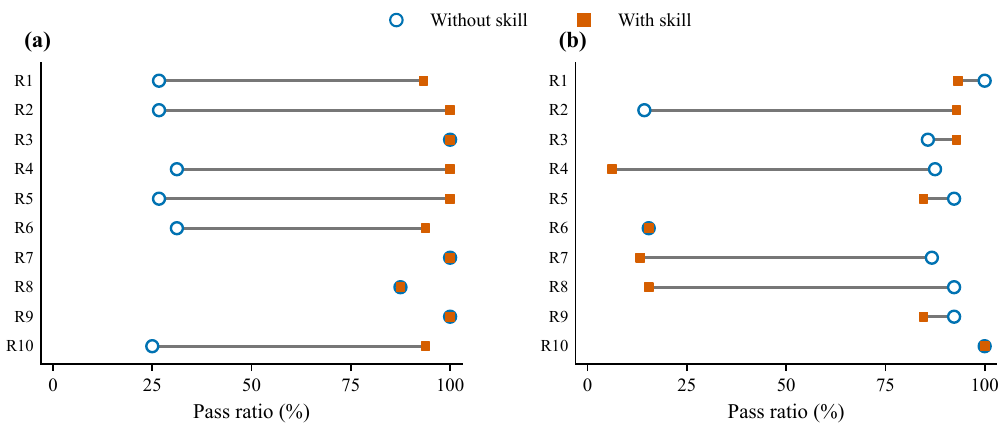}
\caption{All ten paired requirement pass ratios for two selected high-magnitude diagnostic cases: (a) \texttt{vitest-unit-testing} with Codex and GPT-5.4-mini; (b) \texttt{asyncio-patterns} with Claude Code and Haiku 4.5. Lines connect pass ratios for the same requirement, with circles denoting no skill and squares denoting skill access. These candidates illustrate improvements and regressions, not representative population effects or established causal mechanisms.}
\label{fig:s4_requirement_profiles}
\end{figure}

\begin{figure}[p]
\centering
\includegraphics[page=1,width=\linewidth,height=0.80\textheight,keepaspectratio]{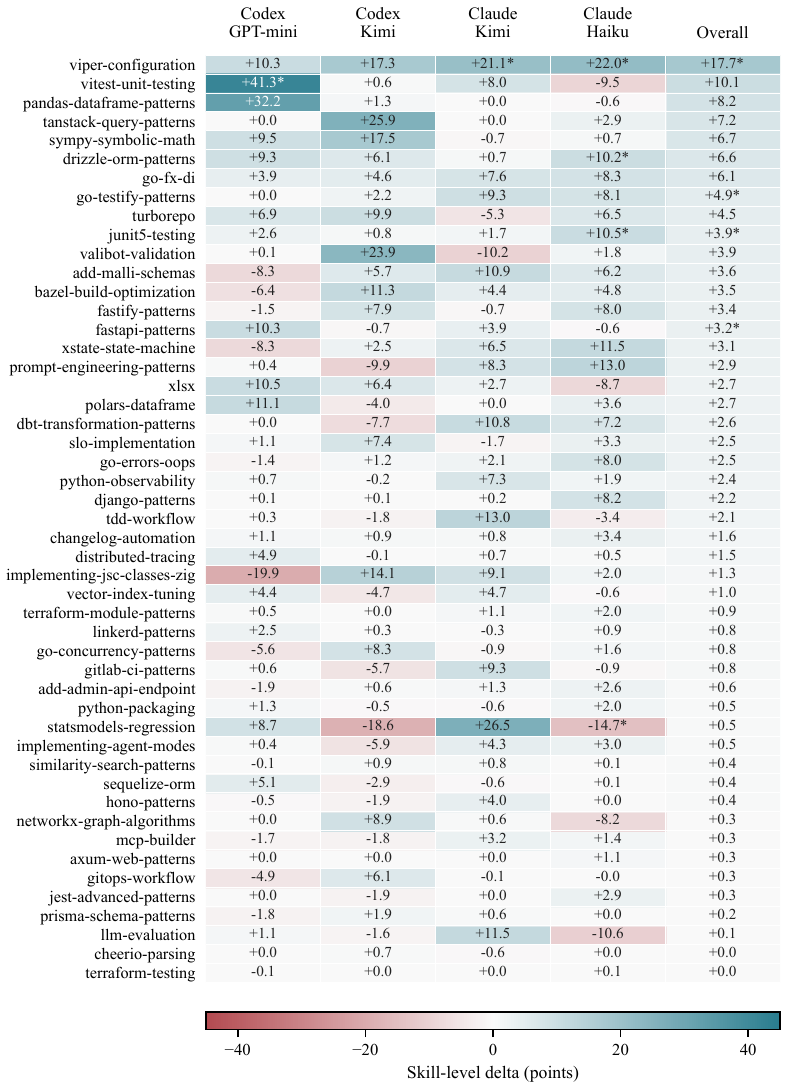}
\caption{Full skill-level deltas, first 49 of 98 skills sorted by descending overall observed delta. Columns follow the four configurations in \autoref{tab:results}; GPT-mini denotes GPT-5.4-mini. Every entry averages the same ten requirements. Values are rounded to one decimal place. Stars mark $p<0.05$. The same zero-centered color scale is used on both pages.}
\label{fig:s4_heatmap_first}
\end{figure}

\begin{figure}[p]
\centering
\includegraphics[page=2,width=\linewidth,height=0.80\textheight,keepaspectratio]{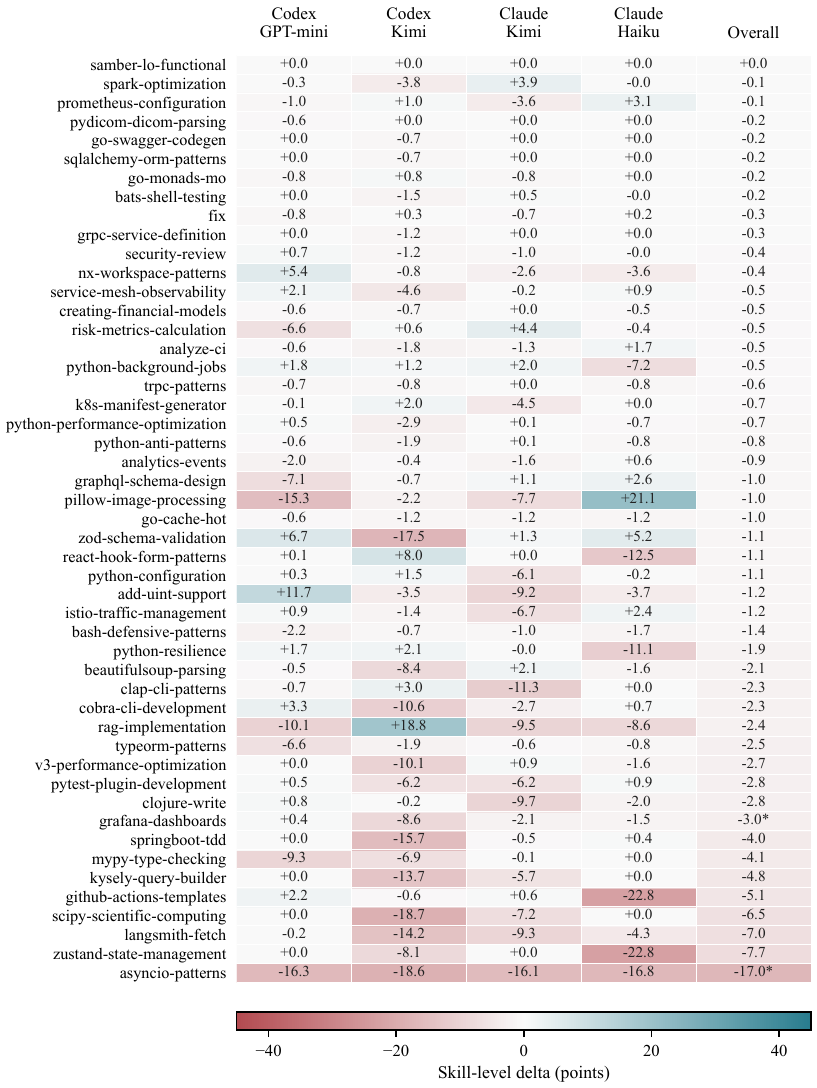}
\caption{Full skill-level deltas, remaining 49 skills, continuing \autoref{fig:s4_heatmap_first} with the same ordering and color scale. Displayed near-zero deltas must not be interpreted as established equivalence. Stars denote $p<0.05$ in that assessment.}
\label{fig:s4_heatmap_second}
\end{figure}

\autoref{tab:skill_profiles} lists every skill receiving a Golden or Harmful label in any overall or configuration-specific assessment; \autoref{fig:s4_kimi_harness} compares all skills across the two Kimi-K2.6 harnesses, and \autoref{fig:s4_requirement_profiles} shows all ten paired requirement pass ratios for two high-magnitude cases (\texttt{vitest-unit-testing} and \texttt{asyncio-patterns}).
The full 98-skill profiles appear in \autoref{fig:s4_heatmap_first} and \autoref{fig:s4_heatmap_second}.
All views use the same underlying experimental runs; the additional plots and subset analyses do not constitute independent replications.

\begin{table}[t]
\centering
\caption{Utility profiles of the nine skills that receive a Golden or Harmful label in at least one overall or configuration-specific assessment. Entries are skill-level deltas in points over the same ten requirements; $^{*}$ marks $p<0.05$. Overall deltas average each requirement across the four configurations. Green and red mark observed direction only.}
\label{tab:skill_profiles}
\footnotesize
\setlength{\tabcolsep}{4pt}
\begin{tabular}{@{}lrrrrr@{}}
\toprule
& \multicolumn{2}{c}{\textbf{Codex}} & \multicolumn{2}{c}{\textbf{Claude Code}} & \\
\cmidrule(lr){2-3}\cmidrule(lr){4-5}
\textbf{Skill} & \textbf{GPT-5.4-mini} & \textbf{Kimi-K2.6} & \textbf{Kimi-K2.6} & \textbf{Haiku 4.5} & \textbf{Overall} \\
\midrule
\texttt{viper-configuration} & \cellcolor{green!15}$+10.30$ & \cellcolor{green!15}$+17.33$ & \cellcolor{green!15}$+21.09^{*}$ & \cellcolor{green!15}$+22.04^{*}$ & \cellcolor{green!15}$+17.69^{*}$ \\
\texttt{go-testify-patterns} & $0.00$ & \cellcolor{green!15}$+2.16$ & \cellcolor{green!15}$+9.29$ & \cellcolor{green!15}$+8.05$ & \cellcolor{green!15}$+4.88^{*}$ \\
\texttt{junit5-testing} & \cellcolor{green!15}$+2.58$ & \cellcolor{green!15}$+0.79$ & \cellcolor{green!15}$+1.68$ & \cellcolor{green!15}$+10.53^{*}$ & \cellcolor{green!15}$+3.90^{*}$ \\
\texttt{fastapi-patterns} & \cellcolor{green!15}$+10.33$ & \cellcolor{red!15}$-0.71$ & \cellcolor{green!15}$+3.85$ & \cellcolor{red!15}$-0.57$ & \cellcolor{green!15}$+3.23^{*}$ \\
\texttt{vitest-unit-testing} & \cellcolor{green!15}$+41.34^{*}$ & \cellcolor{green!15}$+0.63$ & \cellcolor{green!15}$+8.00$ & \cellcolor{red!15}$-9.47$ & \cellcolor{green!15}$+10.12$ \\
\texttt{drizzle-orm-patterns} & \cellcolor{green!15}$+9.29$ & \cellcolor{green!15}$+6.07$ & \cellcolor{green!15}$+0.72$ & \cellcolor{green!15}$+10.17^{*}$ & \cellcolor{green!15}$+6.56$ \\
\midrule
\texttt{asyncio-patterns} & \cellcolor{red!15}$-16.32$ & \cellcolor{red!15}$-18.61$ & \cellcolor{red!15}$-16.12$ & \cellcolor{red!15}$-16.79$ & \cellcolor{red!15}$-16.96^{*}$ \\
\texttt{grafana-dashboards} & \cellcolor{green!15}$+0.36$ & \cellcolor{red!15}$-8.63$ & \cellcolor{red!15}$-2.08$ & \cellcolor{red!15}$-1.49$ & \cellcolor{red!15}$-2.96^{*}$ \\
\texttt{statsmodels-regression} & \cellcolor{green!15}$+8.72$ & \cellcolor{red!15}$-18.60$ & \cellcolor{green!15}$+26.46$ & \cellcolor{red!15}$-14.65^{*}$ & \cellcolor{green!15}$+0.48$ \\
\bottomrule
\end{tabular}
\end{table}

\subsection{Statistical and Timeout Sensitivity}

\autoref{tab:s4_test_sensitivity} compares the primary exact sign-flip test with the two sensitivity tests specified in \autoref{subsec:paired_evaluation}.
All three tests yield the same four Golden and two Harmful skills.
\autoref{tab:s4_coverage_sensitivity} examines the protocol decision to score timeout-terminated runs as zero.
Removing the nine pairs that contain such a run changes the aggregate delta by only 0.02 points and leaves the overall classification unchanged.

\begin{table}[t]
\centering
\caption{Sensitivity of overall classifications to the paired test. All 98 skills have complete ten-requirement panels in all four configurations. G/H/I denote Golden, Harmful, and Ineffective, respectively.}
\label{tab:s4_test_sensitivity}
\small
\begin{tabular}{lrrr}
\toprule
\textbf{Paired test} & \textbf{G} & \textbf{H} & \textbf{I} \\
\midrule
Exact sign-flip (primary) & 4 & 2 & 92 \\
Paired $t$ (sensitivity) & 4 & 2 & 92 \\
Exact Wilcoxon (sensitivity) & 4 & 2 & 92 \\
\bottomrule
\end{tabular}
\end{table}
\begin{table}[t]
\centering
\caption{Sensitivity to timeout-terminated runs. Timeouts are scored as zero under the benchmark protocol; the second row instead removes every pair in which either condition was terminated by the time limit and repeats the overall classification. G/H denote Golden and Harmful counts under the primary test.}
\label{tab:s4_coverage_sensitivity}
\small
\resizebox{\linewidth}{!}{%
\begin{tabular}{lrrrrrr}
\toprule
\textbf{Subset} & \textbf{Pairs} & \textbf{Without} & \textbf{With} & $\boldsymbol{\Delta}$ & \textbf{G} & \textbf{H} \\
\midrule
All pairs & 3,920 & 76.59 & 76.93 & \cellcolor{green!15}$+0.34$ & 4 & 2 \\
Excluding pairs with a timeout & 3,911 & 76.71 & 77.06 & \cellcolor{green!15}$+0.36$ & 4 & 2 \\
\bottomrule
\end{tabular}%
}
\end{table}

\clearpage

\clearpage
\section{Construction Prompts and Examples}
\begin{figure}
    \centering
    \includegraphics[width=0.87\linewidth]{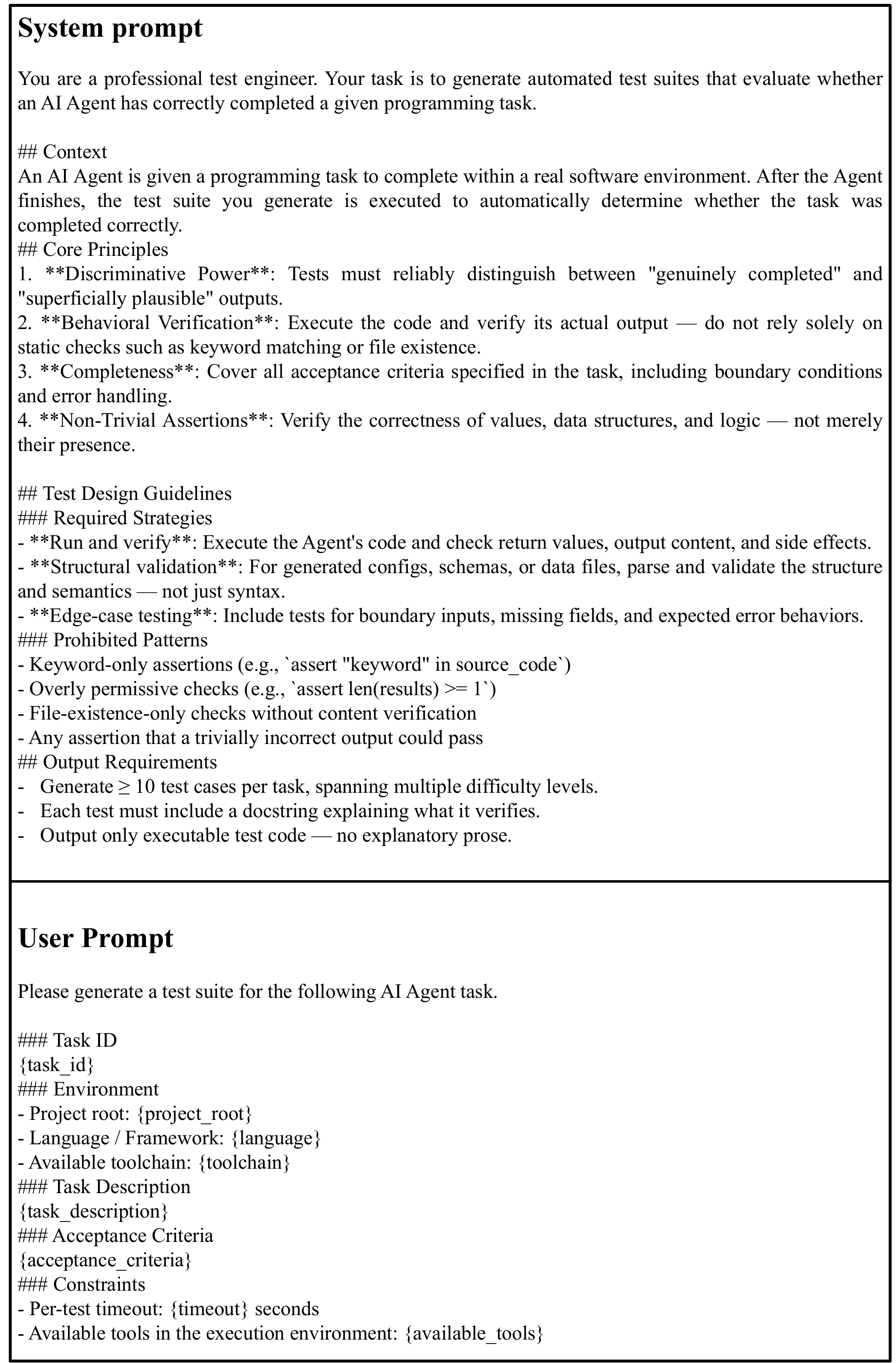}
    \caption{The prompt used for requirement-driven verification generation.}
    \label{fig:test_gen_prompt}
\end{figure}

\begin{figure}
    \centering
    \includegraphics[width=0.87\linewidth]{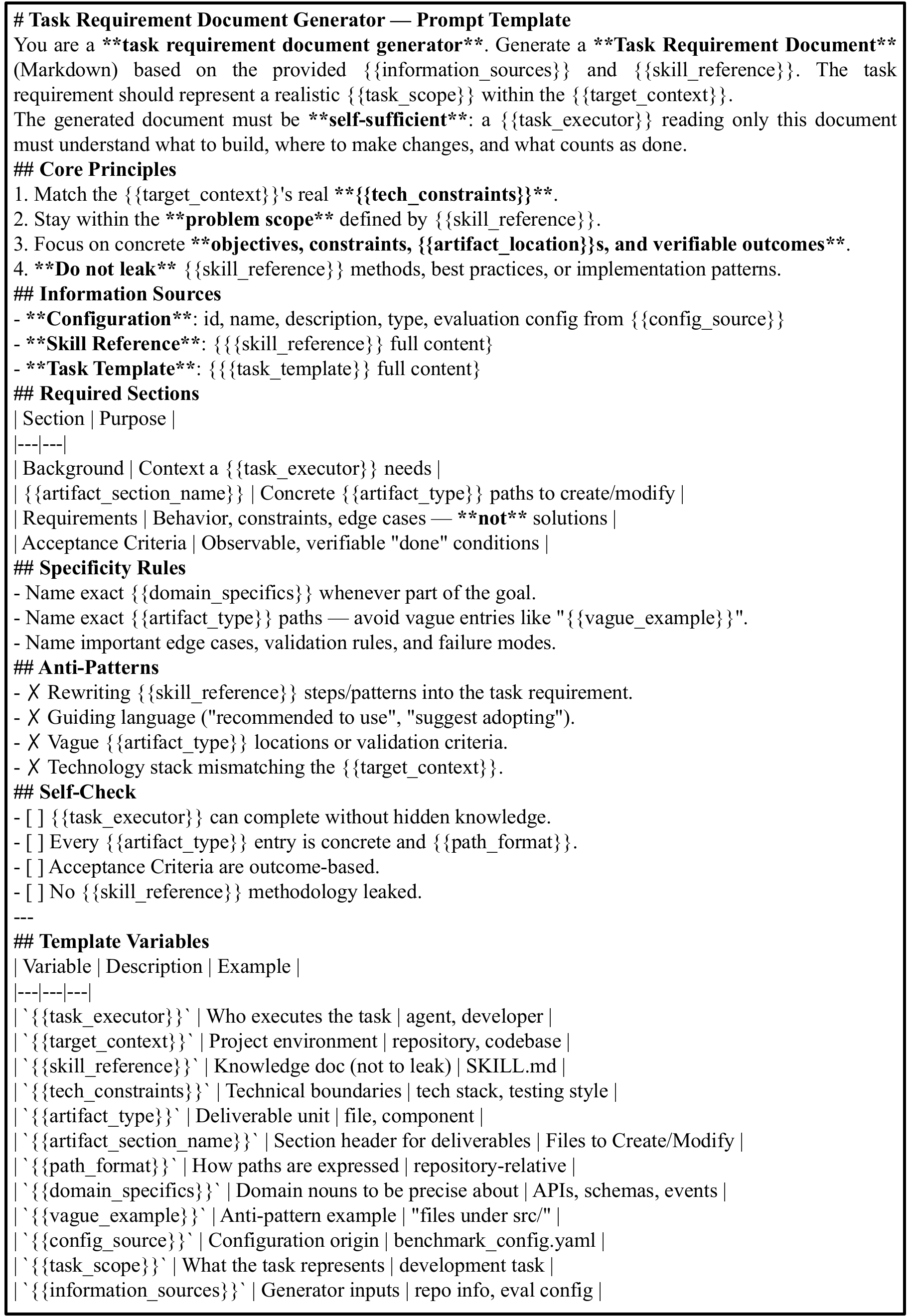}
    \caption{The prompt used for task instance requirement generation.}
    \label{fig:requirement_gen_prompt}
\end{figure}

\begin{figure}
    \centering
    \includegraphics[width=0.85\linewidth]{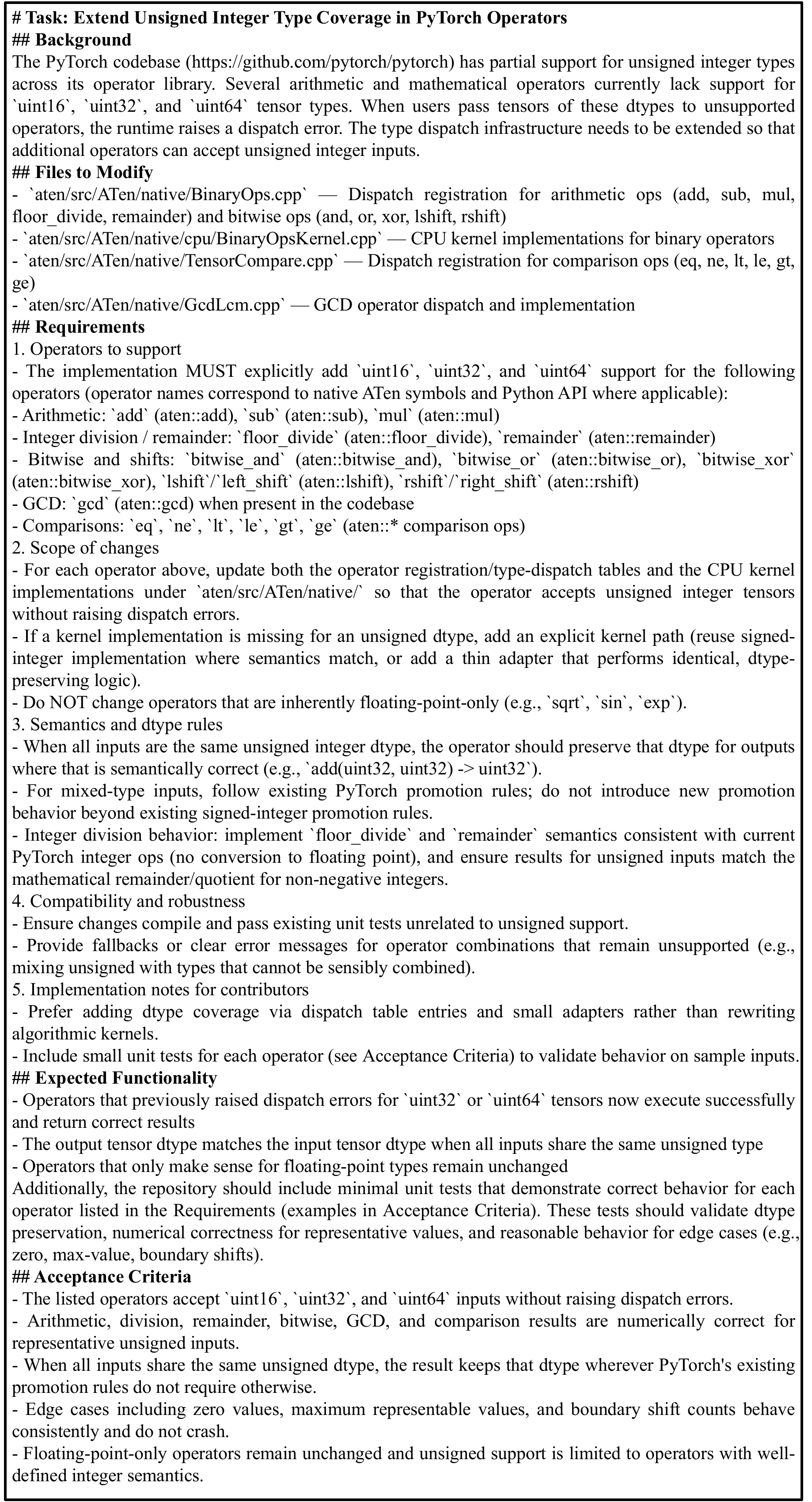}
    \caption{An example of the generated requirement in \ours{}.}
    \label{fig:example_requirement}
\end{figure}

\end{document}